\documentclass[,final
    ,final            % use final for the camera ready runs
  ]
  {aipproc}

\layoutstyle{6x9}

\begin{document}

\title{THE SYNERGY OF GAMMA-RAY BURST DETECTORS IN THE GLAST ERA}



\author{David L. Band, on behalf of the GLAST collaboration}{
address={CRESST and Code 661, NASA/Goddard Space Flight
Center, Greenbelt, MD 20771},
altaddress={CSST, University of Maryland, Baltimore County,
1000 Hilltop Circle, Baltimore, MD 21250}
}

\begin{abstract}
Simultaneous observations by the large number of gamma-ray
burst detectors operating in the GLAST era will provide the
spectra, lightcurves and locations necessary for studying
burst physics and testing the putative relations between
intrinsic burst properties. The detectors' energy band and
the accumulation timescale of their trigger system affect
their sensitivity to hard vs. soft and long vs. short
bursts.  Coordination of the Swift and GLAST observing
plans consistent with Swift's other science objectives
could increase the detection rate of GLAST bursts with
redshifts.
\end{abstract}

\maketitle

%%%%%%%%%%%%%%%%%%%%%%%%%%%%%%%%%%%%%%%%%%%%
%% MAINMATTER
%%%%%%%%%%%%%%%%%%%%%%%%%%%%%%%%%%%%%%%%%%%%

Anticipated to be launched in spring, 2008, the Gamma-ray
Large Area Space Telescope (GLAST) will join a large number
of gamma-ray burst detectors that are already operating in
space.  The strengths of these different detectors
complement each other, both in providing capabilities that
are absent in other detectors and in allowing
cross-calibration.  In this work I compare the different
detectors and their capabilities.

The Table lists burst detectors that will operate during
the first few years of the GLAST mission. Quantitative
comparisons between different missions are difficult
because of the operational details. For example, the
sensitivity usually varies across a detector's
field-of-view (FOV), resulting in a burst detection
threshold that is not uniform.  Because many detectors can
provide spectra over a larger energy band than used for the
burst triggers, I provide two energy bands in the Table.

\begin{table}
\begin{tabular}{lrrrrrrrc}
\hline
    \tablehead{1}{r}{b}{Mission-Detector}
  & \tablehead{1}{r}{b}{Orbit}
  & \tablehead{1}{r}{b}{FOV\tablenote{Field-of-view, in steradians.}}
  & \tablehead{1}{r}{b}{$A_{\rm eff}$ cm}
  & \tablehead{1}{r}{b}{$\sigma$\tablenote{Typical localization uncertainty.}}
  & \tablehead{1}{r}{b}{$\Delta E$ spec.\tablenote{Energy band for spectroscopy.}}
  & \tablehead{1}{r}{b}{$\Delta E$ trig.\tablenote{Energy band for burst trigger.}}
  & \tablehead{1}{r}{b}{$\Delta t$}
  & \tablehead{1}{c}{b}{Ref.}   \\
\hline
GLAST-LAT\tablenote{The LAT FOV is the total sky region from which events
are accepted, and this definition does not account for sensitivity variations
over this area.  The effective area given is the value above $\sim$3~GeV.
The localization depends strongly on the burst intensity and spectrum; a
typical uncertainty for a strong burst is shown.
Both onboard and ground triggers will be used.  The LAT will also observe
GeV band afterglows.}
& 565 km, $\iota$=25.3$^\circ$ & $\sim$3.5 & 8000 & 0.1$^\circ$
& 25 MeV--300 GeV & 25 MeV--300 GeV  &  Variable & \cite{Omodei:2007} \\
GLAST-GBM\tablenote{The FOV is down to the horizon.  The onboard localization
uncertainty is shown; the ground processing will reduce the uncertainty.}
&  565 km, $\iota$=25.3$^\circ$ & $\sim$9 & 122$\times$12 &
8$^\circ$ & 8 keV--30 MeV & 50--300 keV & 0.064--4 s & \cite{Meegan:2007}\\
Swift-BAT\tablenote{The mask open fraction is applied to the effective area.
Note that XRT and UVOT reduce location uncertainties to arcsecond scale.
The XRT and UVOT will observe the optical through X-ray afterglows.}
& 590 km $\iota$=20.1$^\circ$ & $\sim$1.4 & 2600 & 4$^\prime$ &
15--150 keV & 15--150 keV & 0.004--64 s &
\cite{Gehrels:2004,Barthelmy:2005,Band:2006}\\
Konus-Wind\tablenote{Two scintillation detectors pointing in opposite
directions; the sensitivity is low in the plane perpendicular to the
detector axes.}
&  L1 & $\sim4\pi$ & 133$\times2$ & --- & 12 keV--10 MeV
& 45-190 keV & 0.15, 1 s & \cite{Aptekar:1995,Mazets:2004}\\
Suzaku-WAM\tablenote{Four scintillating slabs; the sensitivity is
low in the plane of the slabs.}
& 570 km, $\iota$=31$^\circ$ & $\sim4\pi$ & 800$\times$4 & --- &
50 keV--5 MeV & 110--240 keV & 0.25, 1 s & \cite{Yamaoka:2006} \\
RHESSI &  580 km, $\iota$=38$^\circ$ & $\sim4\pi$ & $\sim150$ & --- &
$>$50 keV & $>$50 keV & Variable & \cite{Wigger:2007} \\
Super-AGILE\tablenote{FOV used is coded in both x and y directions}
& $\sim$580 km, $\iota<3^\circ$ & 1.4 & 312$\times$4 &
1.5$^\prime$ & 15--45 keV & 15--45 keV & Variable
& \cite{Costa:2001,Tavani:2006} \\
AGILE Mini-Cal & $\sim$580 km, $\iota<3^\circ$ & $\sim$2.5 & $\sim$1400 &
---  & 300 keV--100 MeV & 300 keV--100 MeV & Variable
& \cite{Labanti:2006,Tavani:2006} \\
AGILE TKR & $\sim$580 km, $\iota<3^\circ$ & $\sim$2.5 & $\sim$1000 &
15$^\prime$ & 30 MeV--50 GeV & 30 MeV--50 GeV & Variable & \cite{Tavani:2006} \\
INTEGRAL ISGRI/IBIS\tablenote{The effective area includes the mask
opacity.  FOV is within the FWHM.}
& Ecc. & 0.1 & 1300 & 2$^\prime$ & 15 keV--1 MeV
& 15 keV--1 MeV & 8ms--40 s & \cite{Lebrun:2003,Mereghetti:2005}\\
INTEGRAL SPI\tablenote{The effective area includes the mask opacity.}
& Ecc.  & 0.1 & 250 & 10$^\prime$ & 20 keV--8 MeV &
--- & --- & \cite{vonKienlin:2004}\\
INTEGRAL SPI ACS\tablenote{The BGO shields of the SPI cannot
localize bursts.}
& Ecc. & $\sim4\pi$ & $\sim3000$ & --- & --- & --- &
$>50$ ms & \cite{Lebrun:2003,Mereghetti:2005}\\
\hline
\end{tabular}
\caption{Burst Detectors in GLAST Era} \label{tab:a}
\end{table}

Burst triggers ultimately compare an increase in the number
of detected counts in an energy band $\Delta E$ and
accumulation time $\Delta t$ to the expected background
fluctuations; the burst threshold is derived from the
signal-to-noise ratio for a $\Delta E - \Delta t$ bin.  The
burst detection sensitivity is the threshold flux $F_T$
(here over the 1--1000~keV band) as a function of the burst
spectrum (here over 1~s).  Burst spectra can be
parameterized by the `Band' function,\cite{Band:1993}
characterized by low and high energy spectral indices
$\alpha$ and $\beta$, and a characteristic energy $E_p$,
the photon energy of the peak of the $E^2N(E)\propto \nu
f_\nu$. The Figure's left hand panel presents $F_T$ as a
function of $E_p$, fixing $\alpha=-1/2$ and $\beta=-2$, for
different burst detectors. Note that this figure does not
show a detector's sensitivity at a given energy but instead
the detector sensitivity to a burst with a given $E_p$.
Here I show the sensitivity for $\Delta t=1$, but detector
triggers operate with a variety of $\Delta t$ values, and
differ in their sensitivity to bursts with different
durations.\cite{Band:2003,Band:2006}

\begin{figure}
  \includegraphics[height=.25\textheight]{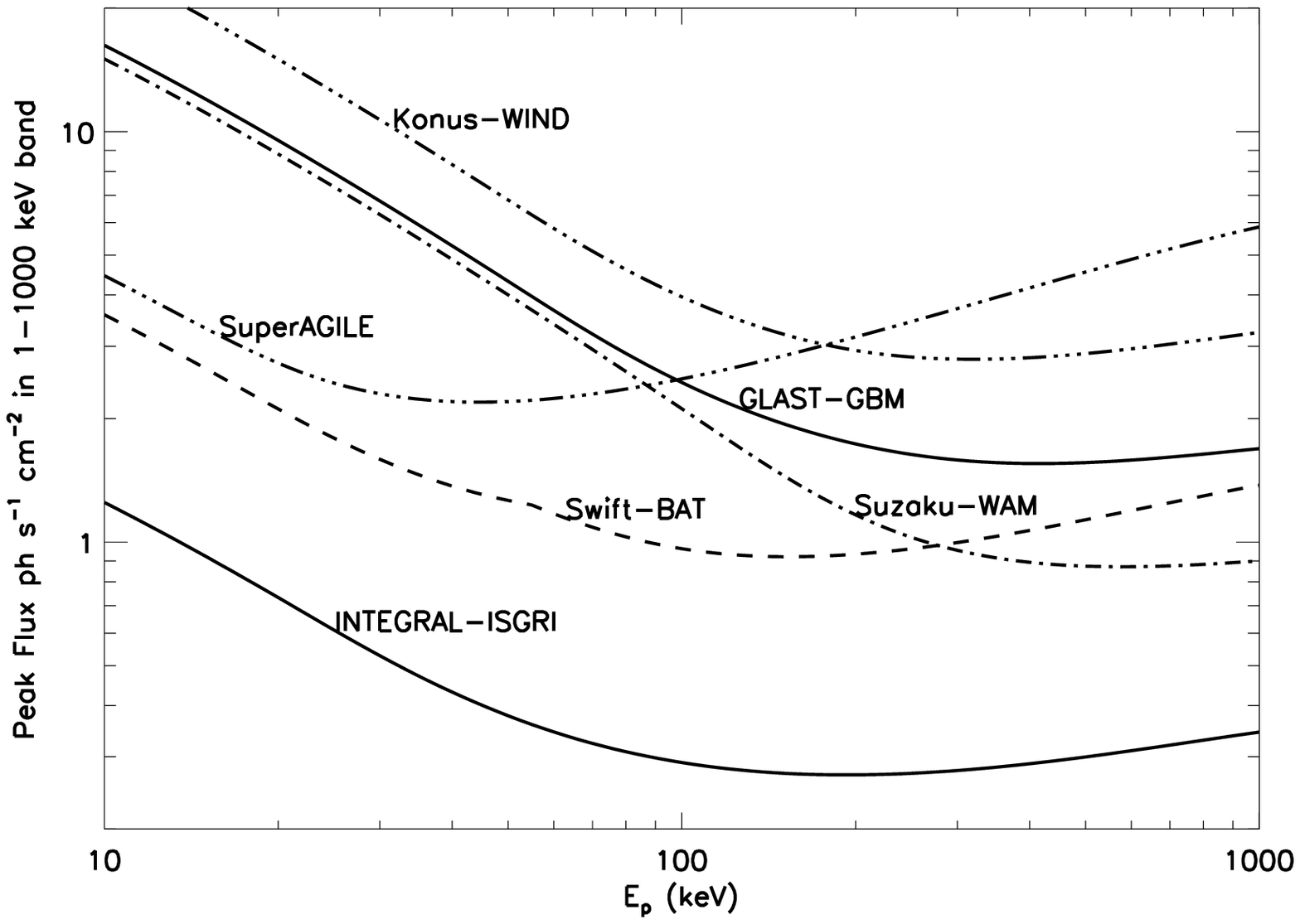}
  \includegraphics[height=.25\textheight]{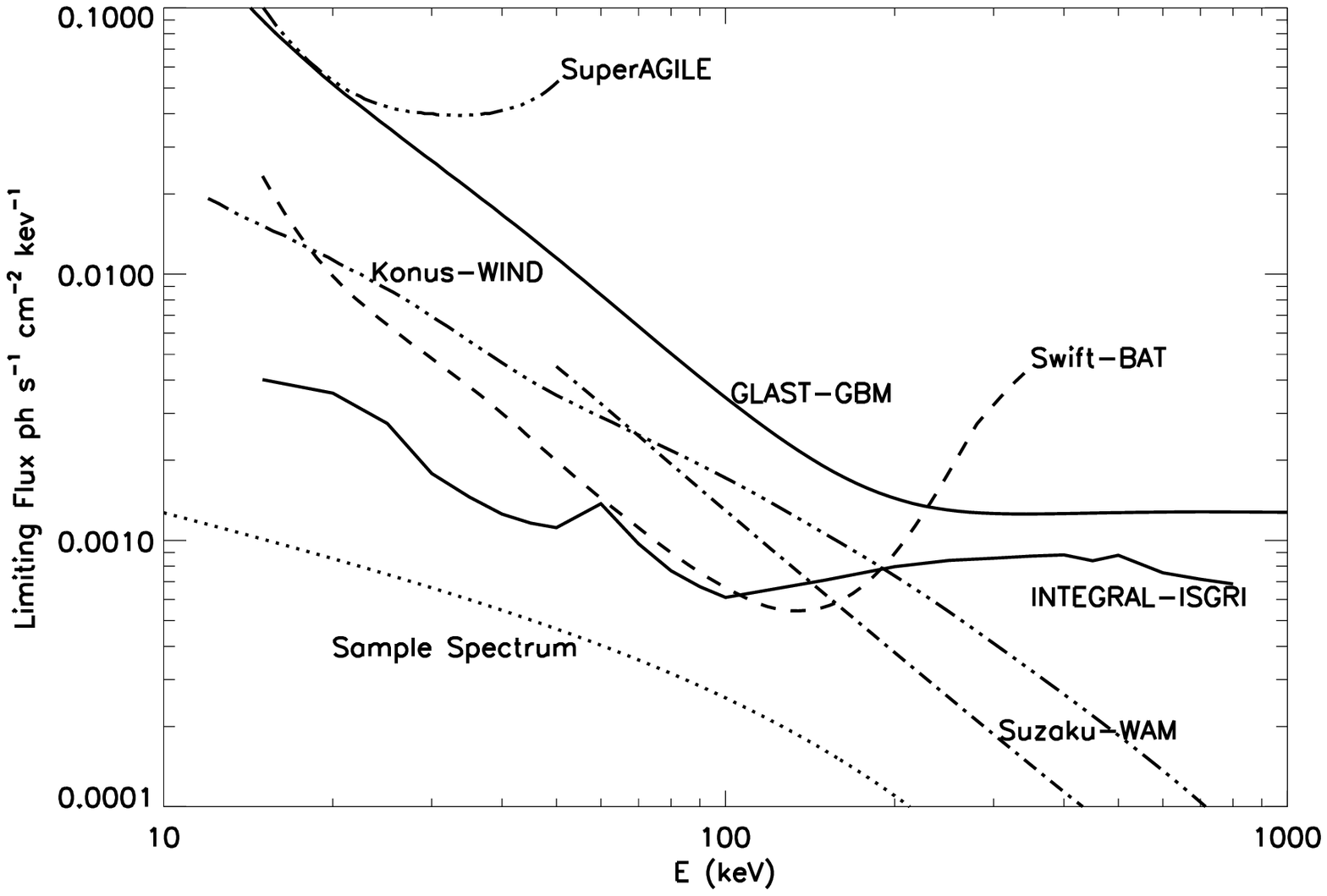}
\caption{Left:  Threshold 1--1000~keV flux as a function of $E_p$ for
different detectors assuming $\alpha=-1/2$, $\beta=-2$ and $\Delta t$=1~s.
Right:  Spectral sensitivity, the flux necessary at $E$ for a 3$\sigma$
measurement in 1~s in a band of width $\Delta E/E=1/2$.}
\end{figure}

In many cases we are not interested in whether a detector
detects a burst---spectral data may be available regardless
of whether the detector triggered---but in the spectra the
detector accumulates. The Figure's right hand panel shows
the detectors' spectral sensitivity, the continuum
sensitivity over a 1~s accumulation time.

The synergy between missions will be maximized by
simultaneous burst observations.  Particularly important is
the overlap between detectors with spectral capability
(GLAST, Konus-Wind, Suzaku-WAM) and localization capability
(Swift-BAT and INTEGRAL-ISGRI). Konus-Wind and INTEGRAL
SPI-ACS essentially see the entire sky, while GLAST-GBM,
Suzaku-WAM and RHESSI see down to the horizon for a
low-Earth orbit. Although ISGRI is sensitive, it has a
(relatively) small FOV.  The GLAST instruments---the LAT
($<$20~MeV--$>$300~GeV) and the GBM (8~keV--30~MeV)---have
large FOVs and the GLAST observatory will operate in a
fixed survey mode. Swift-BAT also has a large FOV and Swift
has a very flexible observing timeline.

Because of their large FOVs and complementary
strengths---localizing bursts and following afterglows for
Swift, accumulating spectra over 7 energy decades for
GLAST---increasing the overlap between these two missions
will have major scientific gains.  During most of its
mission, the LAT's pointing will follow a fixed pattern to
execute an all-sky survey.  On the other hand, Swift
observes a number of targets each orbit with its Narrow
Field Instruments (NFIs)---X-ray Telescope (XRT) and
Ultraviolet-Optical Telescope (UVOT); the wide FOV Burst
Alert Telescope (BAT), which observes bursts' prompt
emission, is centered on the NFIs.  The NFI targets are
burst afterglows and other astrophysically interesting
sources. Semi-analytic calculations show that if Swift does
not coordinate its pointings with GLAST, $\sim$13\% of
GLAST-LAT bursts will be in the Swift-BAT FOV, and
$\sim$27\% of Swift-BAT bursts in the GLAST-LAT FOV.  If
Swift points as close as possible to the LAT pointing
direction these overlap numbers could increase by
$\sim$3$\times$!  However, this estimate neglects Swift's
observational constraints, and sacrifices many of Swift's
scientific objectives. Nonetheless, the judicious choice of
Swift NFI targets could increase the LAT-BAT overlap by
$\sim$2$\times$.  For example, Swift's timeline could
include two sets of targets, one observed when the LAT is
pointed towards the northern hemisphere, and the second for
the southern hemisphere. Procedures to increase this
overlap with little impact on Swift's science objectives
are under development. An increase in the LAT-BAT overlap
would of necessity increase the GBM-BAT overlap.  Note that
a burst in a detector's FOV may nonetheless be too faint to
be detected.

Coordination with GLAST of the timelines of most of the
other burst missions would yield meagre increases in the
detector overlaps because the detectors are nearly all-sky
(or are occulted by the Earth), or their operations do not
lend themselves to such coordination.  Because of the
different FOVs and the varying detection sensitivies, I
forsee that most bursts detected by one of the
constellation of burst detectors will have at best upper
limits from other detectors.  A few bursts a year will be
well-observed by a varying assortment of missions; for
example, Swift might provide a location, Konus-Wind
spectra, and Swift and GLAST-LAT afterglow observations. If
the Swift timeline is coordinated with GLAST, then the LAT
and BAT will observe simultaneously $\sim$50 bursts a year,
although the LAT will probably detect less than half of
these.  The localizations, spectra and lightcurves of this
last set of bursts will advance the study of gamma-ray
bursts.

\begin{theacknowledgments}
I thank members of the BAT team at GSFC and of the GLAST
GRB working group for insightful discussions.
\end{theacknowledgments}

\bibliographystyle{aipproc}   % if natbib is available

\end{document}